\documentclass[11pt]{article}
\pdfoutput=1
\usepackage{authblk}
\usepackage{palatino}
\usepackage{amsmath}
\usepackage{amsthm}
\usepackage{amssymb}
\usepackage{boxedminipage}
\usepackage[mathcal]{euscript}
\usepackage{graphics}
\usepackage{graphicx}
\usepackage{caption}
\usepackage{multirow}
\usepackage{url}
\usepackage{xcolor}
\usepackage{algorithmic}
\usepackage{textcomp}

\newcommand{\calG} {{\cal G}}
\newcommand{\calA} {{\cal A}}
\newcommand{\calX} {{\cal X}}
\newcommand{\calY} {{\cal Y}}
\newcommand{\calZ} {{\cal Z}}

\newcommand{\mypara}[1] {{\bf #1.}}
\newcommand{\wt}[1] {\widetilde{#1}}

\newcommand{\mydeg}[1] {{\tt deg}_{#1}}

\newcommand{\nblk} {{\tt nb}}
\newcommand{\bsize} {{\tt bsize}}

\newcommand{\agg} {{\tt aggregate}}
\newcommand{\rnn} {{\tt RNN}}
\newcommand{\cdgcn} {{\tt CD-GCN}}
\newcommand{\tmgcn} {{\tt TM-GCN}}
\newcommand{\egcn} {{\tt EvolveGCN}}
\newcommand{\eat}[1] {}

\begin{document}
\title{Efficient Scaling of Dynamic Graph Neural Networks
\footnote{A conference version of the paper is to appear in the proceedings of SC'21}
}
\author[1]{Venkatesan T. Chakaravarthy}
\author[1]{Shivmaran S. Pandian}
\author[1]{Saurabh Raje\footnote{The author is currently with the University of Utah}}
\author[1]{\authorcr Yogish Sabharwal} 
\author[2]{Toyotaro Suzumura\footnote{The author is currently with the University of Tokyo}}
\author[2]{Shashanka Ubaru}

\affil[1]{
	IBM Research, India
	\authorcr
	\it{\{vechakra,shivs017,ysabharwal\}@in.ibm.com, saurabh.mraje@gmail.com}
}

\affil[2]{
    IBM T.J. Watson Research Center
	\authorcr
	{\it suzumura@acm.org, shashanka.ubaru@ibm.com}
}
\date{~}

\maketitle     
\begin{abstract}
We present distributed algorithms for training dynamic Graph Neural Networks (GNN) on large scale graphs spanning multi-node, multi-GPU systems. To the best of our knowledge, this is the first scaling study on dynamic GNN. We devise mechanisms for reducing the GPU memory usage and identify two execution time bottlenecks: CPU-GPU data transfer; and communication volume. Exploiting properties of dynamic graphs, we design a graph difference-based strategy to significantly reduce the transfer time. We develop a simple, but effective data distribution technique under which the communication volume remains fixed and linear in the input size, for any number of GPUs. Our experiments using billion-size graphs on a system of 128 GPUs shows that: (i) the distribution scheme achieves up to 30x speedup on 128 GPUs; (ii) the graph-difference technique reduces the transfer time by a factor of up to 4.1x and the overall execution time by up to 40\%. 
\end{abstract}

\section{Introduction}
\label{sec:intro}
Graphs are ubiquitous in diverse domains, ranging from finance to bio-informatics.
Building on classical deep learning, a variety of Graph Neural Networks (GNN) have been
developed for learning graph structured data 
under multiple paradigms such as spectral, convolutional and recurrent GNN \cite{gnn-survey}.

\mypara{Scaling GNN}
Motivated by the success of graph neural networks on real-life learning tasks, 
several recent work have studied the scalability aspects of GNNs. 
PinSage \cite{pinsage} reports an implementation that 
can handle billions of edges.
Ma et al. \cite{neugraph} and Jia et al. \cite{jia} describe efficient 
distributed and multi-GPU implementations.
General purpose GNN libraries, DGL \cite{dgl}, PyG \cite{pyg} and AGL \cite{agl},
and distributed platforms, Aligraph \cite{aligraph} and ${\rm TuX}^2$ \cite{Tux}, 
have been developed. A discussion on software and hardware 
solutions for efficient GNN scaling can be found in the 
survey by Abadal et al. \cite{gnn-hpc-survey}. Recent work by Tripathy et al. \cite{buluc}
presents a detailed study on the data partitioning aspects of GNN scaling.

As part of the above work, various optimization strategies have been developed,
particularly addressing two critical bottlenecks:
GPU memory and communication volume. 
Since the GPU memory is limited when compared to the main memory, 
it is typically infeasible to store large graphs in the GPU in their totality.
Instead, the input graph is transferred in chunks from the CPU to the GPU.
The CPU-to-GPU data transfer affects the overall execution time and prior work (e.g., \cite{neugraph, jia}) 
has designed optimizations based on mechanisms such as data streaming. In large multi-node, multi-GPU systems,
the communication volume is a significant factor in determining the scaling behavior
and different data partitioning methods have been proposed. 
As an example, Aligraph \cite{aligraph} distributes the vertices among
the processors using a hypergraph partitioner and augments it with neighborhood caching.
Tripathy et al. \cite{buluc} argue in favor of multi-dimensional block-wise partitioning 
methods adapted from classical techniques utilized in scaling 
sparse linear algebra.

\mypara{Dynamic GNN}
In many scenarios, graphs are dynamic in nature and evolve over time, e.g., social networks and financial transaction graphs.  
Broadly, two frameworks have been developed to represent dynamic graphs
\cite{dynamic-survey}:
Continuous Time Dynamic Graphs (CTDG) and Discrete Time Dyanamic Graphs (DTDG).
Under the first framework, the evolution of the graph is captured in terms of
insertion/deletion of vertices/edges and updates to the attributes.
The second framework represents the dynamic graph by taking {\em snapshots} at regular intervals
to derive a sequence $G_1, G_2, \ldots, G_T$, where $T$ is the number of timesteps
and $G_t$ is the graph as it stood at timestep $t$.
Various models have been designed for learning within both the CTDG 
(e.g., \cite{Nguyen2018a, Kumar2018b, Trivedi2017, Ma2018b, Rossi})
and the DTDG (e.g., \cite{Chen2018a, Seo2018, cdgcn, Li2017b, egcn, tmgcn}) frameworks.
We refer to the survey by Kazemi et al. \cite{dynamic-survey} for a detailed discussion on the topic.

\mypara{Scaling Dynamic GNN and Our Work}
Our objective is to develop a scalable implementation for training Dynamic GNN models 
on distributed multi-node, multi-GPU systems. 
While the scalability of GNN models (dealing with static graphs) has been well explored,
to the best of our knowledge,  this is the first scaling study for the Dynamic GNN setting.

Our study focuses on the discrete time framework of DTDG.
Dynamic GNN models for DTDG combine GNN from the domain of graph learning and
Recurrent Neural Networks (RNN) from the domain of timeseries analysis.
We consider a generic framework where the model consists of multiple layers,
and each layer  involves 
a  graph convolution component applied over the individual snapshots,
followed by an RNN component applied over the individual vertices across the timeline. 
The former aggregates features from neighboring vertices 
and aids in learning the spatial graph characteristics. The latter captures the temporal aspects.

Multiple dynamic GNN models for DTDG proposed in the literature follow the above framework 
(see survey \cite{dynamic-survey}).
We design optimization strategies catered to the framework and apply them to the three representative models:
{\cdgcn}, {\egcn} and {\tmgcn} \cite{cdgcn, egcn, tmgcn}.
All the three models use the popular Graph Convolutional Network (GCN) \cite{KW} as the GNN component.
Regarding the RNN component, {\cdgcn} employs the popular LSTM \cite{lstm} model,
whereas {\tmgcn} utilizes the M-product \cite{mproduct}.
The {\egcn} model applies LSTM  over the GCN
weight matrices so that the weights evolve by learning the temporal characteristics.
%The models are trained over a certain number of timesteps and then, used to predict properties of future snapshots. 
Prior work has demonstrated the effectiveness of the above models on tasks such as link prediction and node classification.

Strategies developed for scaling static GNN models are also applicable to the dynamic GNN setting.
However, we demonstrate that the timeseries aspect provides specific opportunities, 
which we exploit to design optimization techniques tailored to dynamic GNN. 

{\it Communication Volume:}
A natural data-distribution strategy is to partition the vertices 
and distribute each snapshot according to the above partition among the processors.
Similar to the prior work on GNN, hypergraph partitioners can be utilized to derive an efficient vertex-partitioning.
Under this scheme, the communication volume is dependent on the density properties of the input graph and increases with increase in system size. Furthermore, the communication pattern is highly irregular involving significant implementation overheads,
resulting in poor scaling behavior.

We show that dynamic GNN models allow for an alternative, simple, but effective strategy
based on partitioning the snapshots, instead of vertices.
Under this scheme, the GNN component happens to be communication free
and the RNN component is accommodated via data redistribution.
The salient feature of the approach is that the communication volume is constant at $O(T\cdot N)$ units, 
irrespective of the input graph characteristics and number of processors, where $T$ and $N$ 
are the number of timesteps and vertices, respectively. In contrast to vertex-partitioning based on hypergraphs,
the communication pattern is highly regular with minimal implementation overheads.
This enables efficient scaling to large systems, as demonstrated in our experimental study.

{\it Single Node Optimizations:}
In multi-node systems with multiple GPUs per node,
the architecture  offers significantly higher intra-node CPU-GPU data transfer speeds
among GPUs on the same node, as compared to inter-node communication.
Consequently, the execution time speedup grows sub-linearly with increase in number of nodes,
leading to diminished marginal gains in terms of the ratio of performance to monetary cost.
Hence, it is of interest to consider single node systems (with multiple GPUs) as well.
With the above motivation,  we design two optimization strategies
that are particularly effective on a single node: gradient checkpoint and 
graph-difference based CPU-GPU data transfer.

Gradient Checkpoint:
The GPU memory bottleneck is particularly severe while handling large datasets on a single node. 
For instance,  most of the model-dataset configurations in our experiments do not execute on fewer than $8$ GPUs.
We address the issue by adapting the well-known gradient checkpoint technique \cite{checkpoint}.
Originally designed in the context of deep neural networks with large number of layers,
the method has  been applied to classical RNN models as well \cite{RNN-checkpoint}.
Based on the technique, our implementation 
stores only a subset of snapshots in the GPU at any execution point,
thereby reducing the overall GPU memory usage.

Graph-Difference Based CPU-GPU Data Transfer:
Under the checkpoint-based implementation, the snapshots are not stored permanently in the GPU, 
but get transferred from CPU to GPU on a per-demand basis, leading to increased execution time.
We mitigate the effect based on a crucial observation that, in real world data-sets, the snapshots
evolve at a slow pace and each snapshot is similar in topology (set of edges) to the previous one.
Based on the observation, we design a graph-difference based snapshot transfer method
that offers significant reduction in the transfer time. 

{\it Experimental Evaluation:}
Applying the above strategies, we develop distributed implementations
for the three representative models: {\tmgcn}, {\egcn} and {\cdgcn}.
Our experimental study on a system having $128$ GPUs ($16$ nodes with $8$ GPUs each)
over real-life datasets having up to a billion edges demonstrates that:
(i) the snapshot partitioning scheme enables good scaling behavior and achieves up to $30$x speedup
on $128$ GPUs compared to a single GPU;
(ii) in the singe-node setting,
the graph-difference based strategy offers up to $4.1$x speedup in CPU-GPU transfer time,
resulting in up to $40\%$ reduction in the overall execution time. 
As part of the study, 
we also present a preliminary evaluation comparing snapshot-partitioning and hypergraph-based vertex-partitioning approaches
that demonstrates the better scaling of snapshot-partitioning.

\section{Preliminaries}
\label{sec:prelims}
%
%In this section, we present a brief description of discrete time dynamic graphs and dynamic GNN models.
\subsection{Discrete Time Dynamic Graphs (DTDG)}
A DTDG consists of a dynamic graph $\calG$ and associated input features $\calX$.
The former is a sequence $\calG = G_1, G_2, \ldots, G_T$ over $T$ timesteps, where each $G_t = (V, E_t)$ is a graph,
referred as a {\em snapshot}.
They are defined over the same set of $N$ vertices $V$, but may differ in terms of the spatial topology $E_t$.
Let $A_1, A_2, \ldots, A_T$ be the corresponding sparse adjacency matrices of size $N\times N$,
which can be viewed as sparse tensor $\calA =(A_1, \ldots, A_T)$ of size $T\times N\times N$.
The input features $\calX$ is a sequence $\calX = X_1, X_2, \ldots, X_T$,
where each $X_t$, called a {\em frame},  is a matrix of size $N\times F$ that specifies 
a feature-vector of length $F$ for each vertex in $G_t$.
The sequence $\calX$ can viewed as a dense tensor of size $T\times N\times N$.
Throughout the paper, we use uppercase letters for referring the individual frames/snapshots
and the corresponding calligraphic letters to mean the tensor.

\subsection{Dynamic Graph Neural Networks}
\label{sec:dgnn}
%Dynamic graph neural networks for DTDG combine GNN and RNN models,
%where the former learns the spatial (or graph topological) characteristics,
%and the latter learns the temporal aspects. The two components are discussed below.

\mypara{Graph Neural Networks}
Graph neural networks are meant for learning over static graphs and in our context, they are applied to each snapshot $G_t=(V, E_t)$
in an independent manner. Given the input feature matrix $X_t$ of size $N\times F$,
let $X_t[u]$ denote the feature associated with a vertex $u$.
A GNN model transforms $X_t[u]$ to $Y_t[u]$ via aggregating features of $u$ and its neighors.
Various GNN models have been proposed that differ in terms
of the aggregation operator (see survey \cite{gnn-survey}).
In this paper, we focus on the popular Graph Convolutional Network (GCN) model \cite{KW}
that is employed in the three dynamic GNN models used in our experimental study.

For a vertex $u$, let $\mydeg{u}$ denote the degree of $u$, the number of neighbors. 
Intuitively, to each edge $(u,v)$, the GCN model assigns a weight of $1/\sqrt{(1+\deg_u)\cdot (1+\deg_v)}$.
It derives $Y[u]$ via weighted aggregation over the neighbors and applying a learnable linear layer $W$.
It can be conveniently expressed using the graph Laplacian.

Consider a timestep $t$. Let $A_t$ be the $N\times N$ sparse adjacency matrix of $G_t$.
Let $D$ be the diagonal matrix with $D[u,u] = (1 + \mydeg{u})$ and $I$ be the $N\times N$ identity matrix.
The normalized graph Laplacian is given by:
\begin{eqnarray}
\label{eqn:aaa}
\wt{A} = D^{-1/2}\cdot (A + I)\cdot D^{-1/2},
\end{eqnarray}

The GCN operation is defined as: 
\begin{eqnarray}
\label{eqn:bbb}
Y = \sigma(\wt{A}\cdot X\cdot W),
\end{eqnarray}
where $W$ is a learnable weight matrix of  size $F\times F'$ and $\sigma$ is a suitable activation function such as ReLU.
The output length $F'$ is a tunable parameter.

\mypara{Recurrent Neural Networks}
RNN models are meant for learning over time-series data. In our context, 
they are applied to the time-series corresponding to each vertex in an independent manner.
Given a vertex $u$ with features $X[u] = X_1[u], X_2[u], \ldots, X_T[u]$ each of length $F$, the RNN model produces a transformed
series $Y[u] = Y_1[u], Y_2[u], \ldots, Y_T[u]$ each of length $F'$, a tunable parameter.
For each timestep $t$, the model maintains a hidden state $S_t[u]$. 
It generates $Y_t[u]$ and $S_t[u]$ by considering the previous state $S_{t-1}[u]$, 
input features from previous and current timesteps, the new features from previous steps:
%
%\begin{eqnarray*}
%(Y_t[u], S_t[u]) \ = \ \rnn(S_{t-1}[u], \ X_{t-w}[u],..X_t[u], \ Y_{t-w}[u],..Y_{t-1}[u]),
%\end{eqnarray*}
\begin{eqnarray*}
(Y_t[u], S_t[u]) 
&=& \rnn(S_{t-1}[u], \\
&\quad& \quad \quad X_{t-w}[u],\ldots, X_t[u], \\
&\quad& \quad \quad Y_{t-w}[u], \ldots Y_{t-1}[u]),
\end{eqnarray*}
where the parameter $w$ controls the prior window length.
The RNN may involve internal learnable parameters.
Taken over all the vertices, the RNN operation can be expressed as:
\begin{eqnarray}
\label{eqn:rnn}
(Y_t, S_t) 
&=& 
\rnn(S_{t-1}, 
\quad X_{t-w},\ldots, X_t, 
\quad Y_{t-w}, \ldots Y_{t-1}),
\end{eqnarray}
%\begin{eqnarray}
%\nonumber
%(Y_t, S_t) 
%&=& \rnn(S_{t-1}, \\
%\nonumber
%&\quad& \quad X_{t-w},\ldots, X_t, \\
%\label{eqn:rnn}
%&\quad& \quad 
%Y_{t-w}, \ldots Y_{t-1}),
%\end{eqnarray}
wherein $Y_j$ is of size $N\times F'$ and $S_j$ are of size $N\times s$, with $s$ being a tunable RNN hidden state length.

Different RNN models have been proposed in the literature.
Among the dynamic GNN models used in our study, {\cdgcn} and {\egcn} employ LSTM \cite{lstm},
whereas {\tmgcn} is based on the M-product \cite{mproduct}. We shall describe the two RNN models,
while discussing the above dynamic GNN models later in the paper.

\begin{figure}
\centering
\begin{tabular}{ccc}
\includegraphics[width=2in]{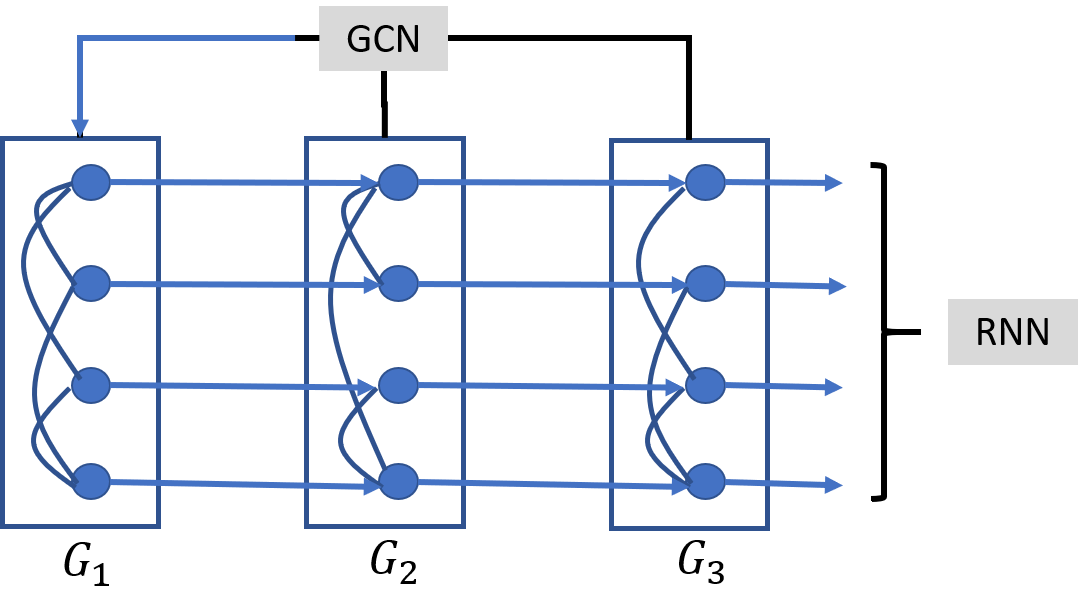}
&
\quad
&
\includegraphics[width=3.3in]{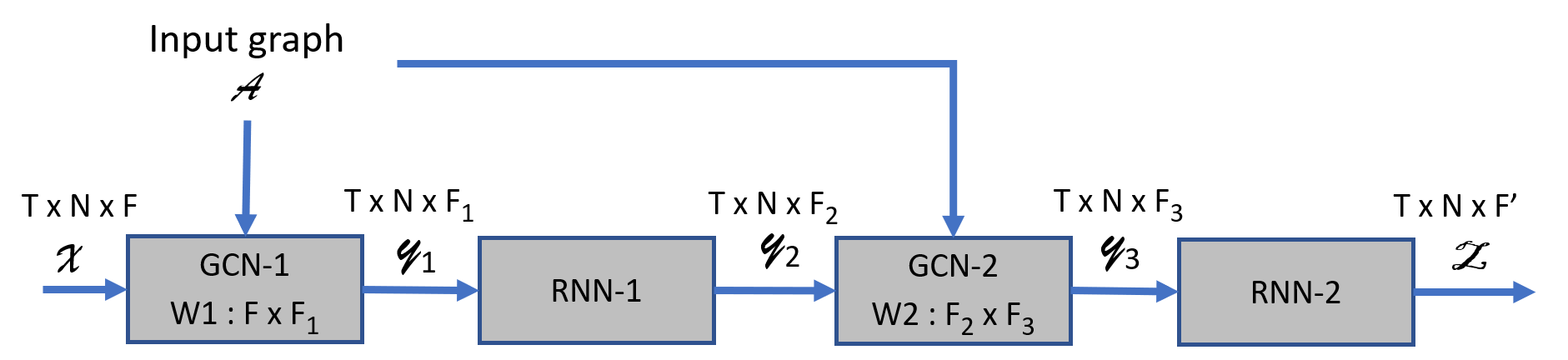}
\\
(a) GCN and RNN operations
&
\quad
&
(b) Multi-layer dynamic GNN
\end{tabular}
\caption{
Dynamic GNN. Part (a) illustrates a single pair of GCN and RNN operations over 
$N=4$ vertices and $T=3$ timesteps. 
Part (b) presents a two layer model with each layer consisting of a GCN-RNN pair.
}
\label{fig:dgnn}
\end{figure}

\mypara{Dynamic Graph Neural Networks for DTDG}
Our work applies to a family of dynamic GNN models for DTDGs,
which we abstract using the framework described below.
Details specific to the three representative models ({\tmgcn}, {\cdgcn}, {\egcn})
used in our experimental evaluation are described in Section \ref{sec:models}.

Under the framework, a dynamic GNN model consists of multiple layers.
Each layer involves a GCN \cite{KW} module
operating on each snapshot independently, followed by an RNN module 
operating on the feature-vector of each vertex independently along the timeline. 
Figure \ref{fig:dgnn} (a) illustrates the idea.

A multi-layer model is constructed by iterating over this pair of GCN/RNN operations. 
Figure \ref{fig:dgnn} (b) illustrates a two-layer model.
Here,  the input dynamic graph is represented as a sparse tensor $\calA$ of size $T\times N \times N$, 
and the input features $\calX$ is a dense tensor of size $T\times N \times F$. 
Tensor notation is used to represent the intermediate activations as well. 
While the two RNN components operate on the 
intermediate features output by the previous module, the GCN components apply graph convolution on the input dynamic graph.

The intermediate feature lengths $F_1, F_2, F_3$, and the embedding length $F'$ are tunable.
They   determine the size of the GCN weight matrices and the internal parameters of the RNN.
The output of the iterative process is a tensor $\calZ$  of size $T\times N\times F'$ 
that provides an embedding of size $F'$ for each $u$ at each timestep $t$.

The embeddings can be used in multiple ways. In vertex classification, we are given ground truth labels for each 
vertex at each timestep in the form of a matrix 
$Q$ of size $T\times N$ with entries from $\{1, \ldots, C\}$, where $C$ is the number of categories. 
For this application, we derive predictions by projecting each embedding matrix $Z_t$ to the label space
via a learnable weight matrix $U$ of size $F' \times C$. The predictions are compared against the ground truth using a 
loss function such as cross-entropy. Edge prediction can be performed via concatenating the embeddings of 
the edge end-points. The latter is explored in our experimental evaluation.

The weight matrices associated with the GCN and the RNN modules, and the matrix $U$ are the learnable parameters of the model.
Backpropagation of the gradients is performed after the forward phase has completed the processing
of all the snapshots (all their vertices).

\begin{figure}
\centering
\includegraphics[width=5in]{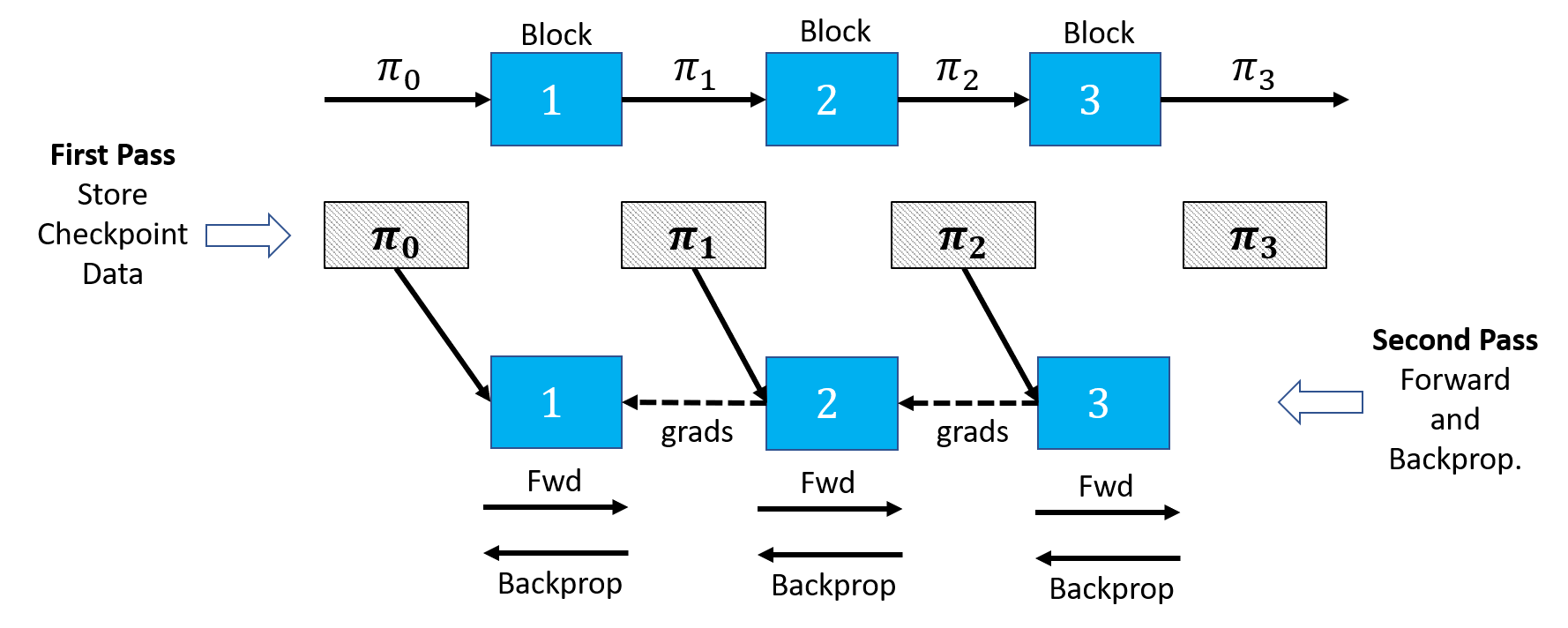}
\caption{
Gradient checkpoint illustration. Here, number of blocks $\nblk=3$. $\pi_j$ represents the RNN-specific data passed
from block $j$ to $j+1$, which gets stored as part of checkpointing. $\pi_0$ is the initial data.}
\label{fig:checkpoint}
\end{figure}

\section{Single GPU Implementation}
In this section, we discuss optimizations of gradient checkpoint and graph-difference based CPU-GPU transfer.
\subsection{Gradient Checkpoint}
\label{sec:checkpoint}
The standard two-phase training process, consisting of forward and backpropagation, involves 
storing a copy of the intermediate activation tensors, as well as the inputs $\calA$ and $\calX$.
This leads to severe GPU memory bottleneck for large inputs.
In our experiments, most of the model-dataset configurations do not execute on fewer than $8$ GPUs.
We adapt the well-known gradient checkpoint method (e.g., \cite{checkpoint, RNN-checkpoint})
to optimize the memory requirements.

In our setting, the GCN component operates independently on each snapshot, but inter-dependency is caused by the RNN component
acting along the timeline. We partition the timeline into $\nblk$ blocks, each containing $\bsize = T/\nblk$ timesteps,
where the number of blocks $\nblk$ is a tunable parameter.
For a block $b\in [1, \nblk]$, the range of timesteps is given by the 
starting and ending timesteps $s(b) = 1 + (b-1)\cdot \bsize$ and $e(b) = b \cdot \bsize$.

The idea of checkpointing is to restrict the storage of the input and the intermediate data to a single block at any point during the execution.
Towards that goal, we first execute the forward pass in the usual manner by processing the blocks in the increasing order.
Then, the backpropagation pass is conducted in the reverse order, starting with the last block.
The processing of each block $b$ consists of two parts: a rerun of the forward pass,
followed by gradient propagation in the reverse direction. 
The process limits the memory usage to a single block thereby reducing the overall the memory requirement. 
See Figure \ref{fig:checkpoint} for an illustration.

The procedure requires the ability to re-execute a block $b$. 
The GCN component does not have dependency on the prior block $b-1$. 
However, the RNN component requires the following data computed in block $b-1$:
(i) the RNN hidden state corresponding to the last timestep of block $b-1$ (namely $S_{e(b-1)}$); 
(ii) the activations of the RNN corresponding to the last $w$ timesteps of the block $b-1$,  
where $w$ is the window size (see Eqn. \ref{eqn:rnn}).
We denote the above information passed from block $b-1$ to block $b$ as $\pi_{b-1}$ (see Figure \ref{fig:checkpoint}).
We store $\pi_b$ for all the blocks during the forward pass, to be reused during backpropagation.

The total GPU memory requirement involves two components: memory needed to store the activations of the current block 
and the checkpoint data. 
The former consists of the snapshots $A_{s(b)}, \ldots,$\break $A_{e(b)}$, input features $X_{s(b)}, \ldots, X_{e(b)}$, 
and the intermediate tensors. The latter consists of the checkpoint data $\pi_b$ stored across all the blocks. 
While the former intra-block memory requirement is determined by the 
block size $\bsize=T/\nblk$, the checkpoint data is determined by the number of blocks $\nblk$. 
The two components can be balanced by adjusting the parameter $\nblk$.

The parameter $\nblk$ not only determines GPU memory usage, but also influences the execution time,
since the GPU utilization is better and the latency is lower under larger block sizes (fewer blocks).
In our experiments, we tune the parameter so as to achieve the best possible execution time,
while ensuring that the GPU memory usage does not exceed the available memory.

\subsection{Graph-difference Based Input Transfer}
\label{sec:gd}
In order to save memory, our checkpoint implementation
stores only the input and the intermediate data corresponding the current block $b$ in the GPU. 
While latter gets generated and resides on the GPU,
the input comprising of the snapshots and the features corresponding to the block $b$
get transferred from the CPU to the GPU.
This transfer happens twice, once during the forward phase and the second during the rerun segment of the backpropagation.
We use pinned memory to optimize the above data transfer, as this avoids the use of paged memory. 
In spite of the optimization,  our experiments show that the transfer time
constitutes an important component of the overall execution. 
In this section, we exploit the properties of dynamic graphs to devise a graph-difference based method that reduces
the transfer time.

Our method is motivated by the fact that dynamic graphs change gradually and therefore
consecutive snapshots are expected to have substantial overlaps in their topology.
In addition, as explained later (c.f. Section \ref{sec:models}),  towards improving accuracy,
{\tmgcn} and {\egcn} apply certain  pre-processing steps, named M-product and edge-life. 
These steps tend to smoothen the differences across the snapshots,
and as a result, they magnify the overlaps in the topology among consecutive snapshots. 

Consider a block $b$ pertaining to the sequence of snapshots\break $A_{s(b)}, \ldots, A_{e(b)}$ of length $\bsize$.
The first snapshot $A_{s(b)}$ is transferred from the CPU to the GPU using standard 
sparse matrix representation of (index,value) pairs.
Consider two consecutive snapshots $A_i$ and $A_{i+1}$. Assuming that $A_i$ is already present in GPU,
we describe how the graph-difference method transfers $A_{i+1}$. 

We partition the edges of $A_i$ and $A_{i+1}$ into three sets: 
\begin{itemize}
\item $A^{com}$: the set of common edges present in both $A_i$ and $A_{i+1}$,
\item $A_i^{ext}$: the extra edges present in $A_i$ but not in $A_{i+1}$, and
\item $A_{i+1}^{ext}$: the extra edges present in $A_{i+1}$ but not in $A_i$.
\end{itemize}
Now, instead of transferring $A_{i+1}$ using standard sparse matrix representation,
we only transfer:
\begin{itemize}
\item the indices corresponding to $A_i^{ext}$ 
\item the indices corresponding to $A_{i+1}^{ext}$ 
\item all the values for the new snapshot $A_{i+1}$
\end{itemize}
We first derive the common indices $A^{com}$ by excluding $A_i^{ext}$ from $A_i$.
We then reconstruct the indices of the new snapshot $A_{i+1}$
by adding  the extra edges in $A_{i+1}^{ext}$ to $A^{com}$.
While the snapshots overlap in terms of the topology, the values associated with their
edges are not expected to overlap. So, the transfer of value of the new snapshot is required.
When there is a large overlap in the structure, this results in substantial saving as it avoids
transferring the indices for the common structure of the snapshots $A_i$ and $A_{i+1}$.

\section{Distributed Implementation}
In the multi-node setting, the communication volume is a critical aspect and it is determined by the data partitioning.
We first discuss a vertex partitioning approach, adapted from the static GNN setting,
and then present our snapshot partitioning approach.
Assume that we have $P$ processors, each endowed with a GPU, which could be cores of the  same node or span multiple nodes. 

\subsection{Vertex Partitioning Approach}  
\label{sec:vertex}
A common approach used in (static) GNN setting with a single input graph
is to partition the vertices among the processors (e.g, \cite{aligraph}). 
Adapting to our setting, we partition the vertex set $V$ uniformly among the processors so that 
each processor $p$ owns $N/P$ vertices, denoted $V_p$. 
The snapshots get partitioned accordingly:
for each $t$, the rows of $A_t$ corresponding to $V_p$ are stored at processor $p$. 
Each input feature matrix $X_t$ is partitioned in a similar manner.

The RNN component operates independently on each vertex.
Hence, each processor $p$ can perform the operation on the set of vertices $V_p$
without having to communicate with the other processors.
Thus, the RNN component is communication free. 
However, the GCN component requires significant communication.

Consider the GCN operation on a snapshot $A_t$. Each vertex $u$ aggregate
the neighborhood features. Viewed from the other direction, the feature of a vertex $v$ 
is required by all its neighbors $\Gamma_t(v)$, which may be distributed among multiple processors.
Let $\lambda_t(v)$ denote the number of processors that own at least one neighbor of $v$.
Then, the communication is $\lambda_t(v)$ units. Summed across all snapshots and vertices,
the total communication is given by $\sum_t \sum_v \lambda_t(v)$ units per GCN module,
where a unit refers to a feature vector.
The partition minimizing the above communication volume can be found using 
hypergraph partitioners such as PaToH \cite{patoh}.

\mypara{Shortcomings of Vertex-partitioning}
Under vertex partitioning, 
the communication volume increases with the number of processors $P$ and is dependent on the graph density.
In addition, the communication pattern is irregular, resulting in significant implementation overheads.
Finally, the approach requires sophisticated hypergraph partitioners that incur high preprocessing time.
Similar observations have been made in recent prior work on scaling GNN \cite{buluc}.
The dynamic GNN allows us to design simpler and effective partitioning algorithm that overcomes the above issues.

\subsection{Snapshot Partitioning and Redistribution}
\label{sec:dist}
The core idea of our scheme is to partition the snapshots among the processors, instead of the vertices.
We then accommodate the RNN component via a re-distribution of the feature matrices.

\begin{figure}
\centering
\begin{tabular}{c}
\includegraphics[width=6in]{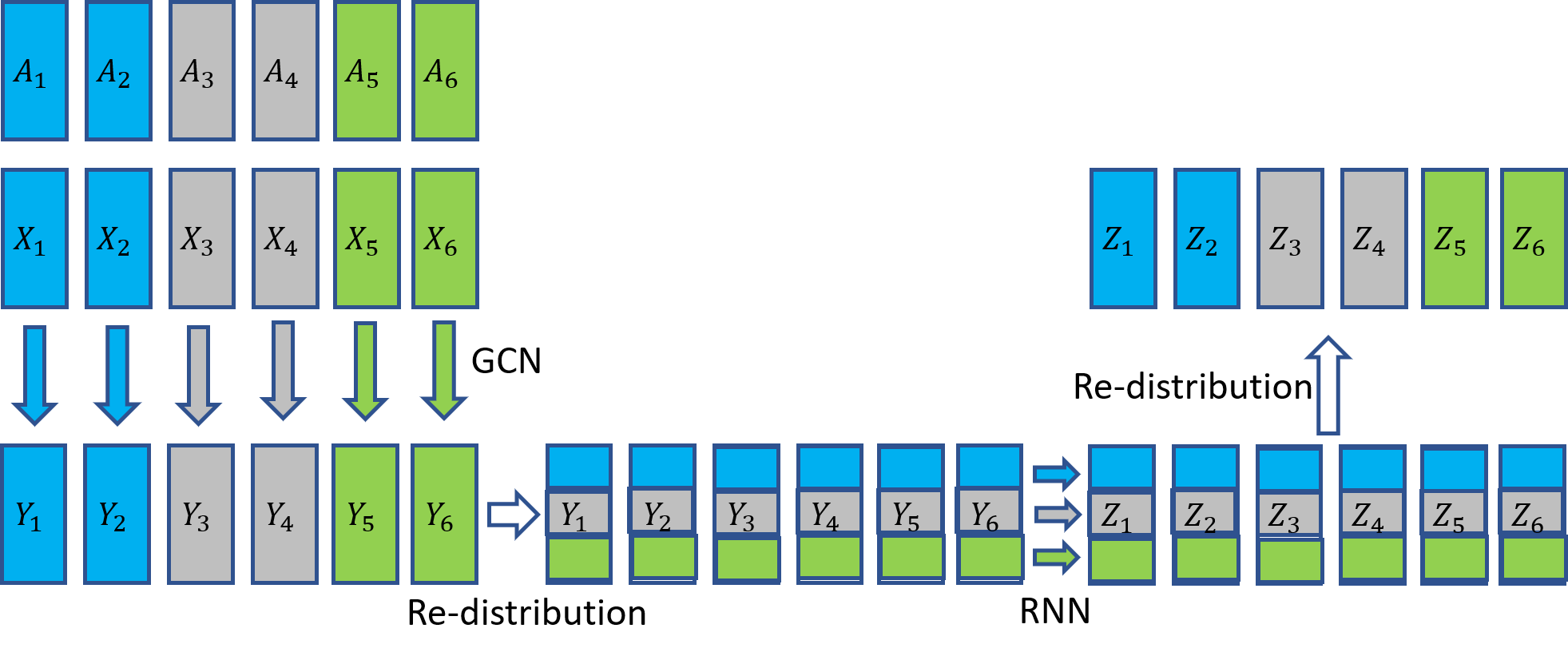}\\
(a) Non-checkpoint setting\\
\includegraphics[width=6in]{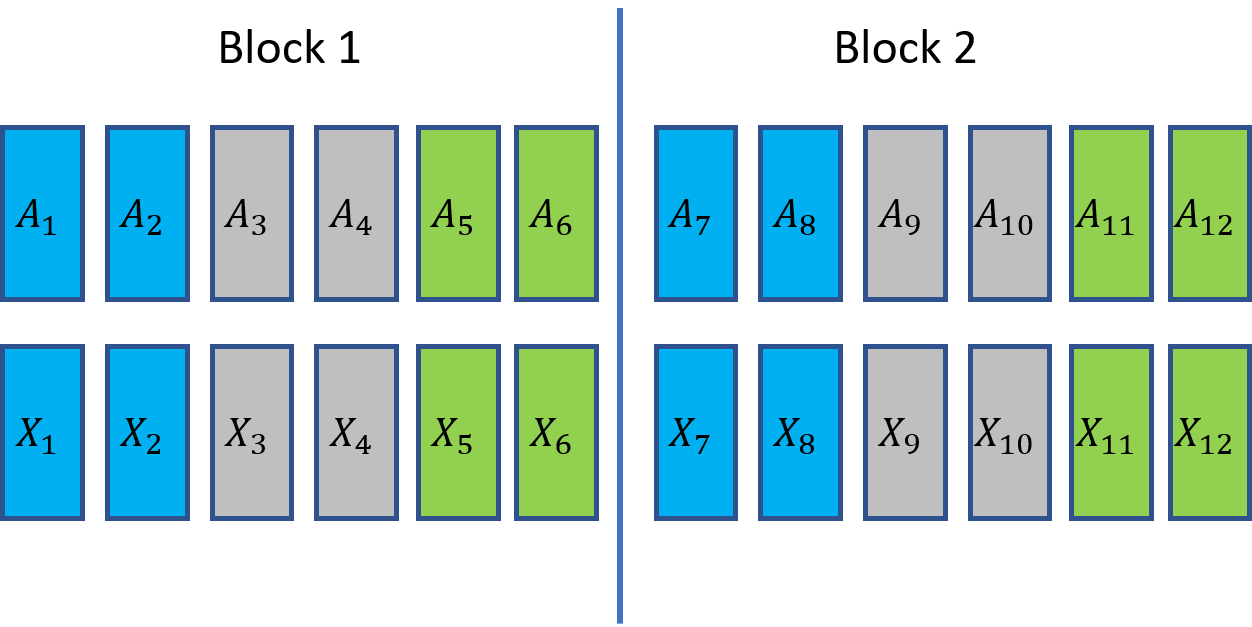}\\
(b) Checkpoint setting
\end{tabular}
\caption{
Snapshot partitioning and re-distribution. Part (a) illustrates the process without checkpoint taking $T=6$ timesteps
and  $P=3$ processors, represented by the three colors. The matrices $A_t$ are sparse and are of size $N\times N$, whereas 
the other matrices are feature matrices of size $N\times F$ (with different feature lengths $F$). 
The figure shows the partitioning of the snapshots $A_t$  and the input features $X_t$, as well the two re-distributions and 
the GCN/RNN operations.  Part (b) illustrates the partitioning in the checkpoint setting taking $T=12$, $P=3$ and the number of  blocks $\nblk=2$.}
\label{fig:dist}
\end{figure}

\mypara{Snapshot Partitioning and GCN} 
For the ease of exposition, we first discuss the implementation without gradient checkpoint. Consider the first layer of the model
involving a pair of GNN and RNN components.
Figure \ref{fig:dist} (a) illustrates the partitioning and the execution of the GCN/RNN components described below.

We partition the snapshots among the processors in a contiguous manner so that each processor 
owns $k=T/P$ contiguous snapshots. Namely, processor $p$ is assigned snapshots $A_s$ to $A_e$, 
where $s= 1 + (p-1)\cdot k$ and $e=p \cdot k$. Similarly, the input features $X_s$ to $X_e$
are assigned to $p$. The GCN weight matrices $W$ are very small in size  and  we store a copy of the matrices in all the processors.

For each $t$, the processor responsible for the timestep $t$ has both $A_t$ and $X_t$ in entirety,
and so it can perform the GCN operation $Y_t=\wt{A}\cdot X_t\cdot W$ by itself without communication.
Thus, the GCN component is communication free. 
Let $Y_1, Y_2, \ldots, Y_T$ denote the output matrices of the GCN operation, where $Y_t$ is generated at the processor responsible for the timestep $t$. 

\mypara{Re-distribution and RNN}
The RNN module is applied over the sequence $Y_1, Y_2, \ldots, Y_T$. 
The module operates on each vertex $u$ independently and requires the 
entire sequence \break $Y_1[u], Y_2[u], \ldots, Y_T[u]$, due to dependency across the timeline. 
To facilitate the process, we re-distribute the matrices by performing a vertex-level partitioning. 
We partition the vertex set $V=\{v_1, v_2, \ldots, v_N\}$ into $P$ chunks of size $k=N/P$ each
and make processor $q$ the owner of the $q^{th}$ chunk.
Namely, the processor $q$ owns the vertices $V_q = \{v_s, \ldots, v_e\}$,  
where $s=1 + (q-1)\cdot k$ and $e=q\cdot k$. 
 
For each timestep $t$, the processor $p$ responsible for the timestep splits the matrix
$Y_t$ into $P$ chunks and sends the $q^{th}$ chunk to the processor $q$.
The processor $q$ assembles the sequence \break $Y_1[V_q], Y_2[V_q], \ldots, Y_T[V_q]$
and applies the RNN operation. The data transfers are realized via an all-to-all communication.

Let the output of the operation be $Z_1[V_q], \ldots, Z_T[V_q]$.
The dynamic GNN model may involve multiple layers. To prepare for the GCN model at the next layer,
we  re-distribute the $Z$ matrices to match the original snapshot partitioning. 
Namely, for each $q$ and $t$, the processor $q$ sends $Z_t[V_q]$ to the processor $p$ responsible for timestep $t$.
Upon receiving the data, each processor $p$ can reassemble the matrix $Z_t$ for each timestep it is responsible for.
As before, the data transfers are realized via an all-to-all communication.

\mypara{Gradient Checkpoint Implementation}
We next adapt the partitioning algorithm to the context of gradient checkpoint.
Assume that we have $\nblk$ blocks each having $\bsize=T/\nblk$ timesteps.
We apply snapshot partitioning within each block so that each processor
is responsible for $\bsize/P$ timesteps within the block.
Consequently, snapshots assigned to a processor are contiguous within a block,
but non-contiguous when viewed over the entire timeline.
See Figure \ref{fig:dist} (b) for an illustration.

The above block-wise partitioning facilitates the RNN computation.
The processors operate within the same block and move to the next in a synchronous fashion.
For each block, the GCN operations are applied over the timesteps in the block
and the RNN operation is executed restricted to the block. Similarly,
the all-to-all communication are also limited to feature matrices of the block.
Finally, checkpoint data is stored and the procedure advances to the next block.

\mypara{Communication Volume}
For every dynamic GNN layer consisting of a GCN-RNN pair, we perform two re-distributions.
Each involves an all-to-all communication with an overall volume of $T\cdot N$ units,
where a unit refers to a feature vector. 
Regarding backpropagation, at a high level,  the procedure is executed in a symmetrically opposite manner
via performing the above steps in the reverse order. Akin the to the forward phase, the procedure involves 
two gradient re-distributions, realized via all-to-all communications.
Thus, the overall communication volume is $O(T\cdot N)$ units.

\mypara{Advantages of Snapshot Partitioning}
An important benefit of snapshot partitioning is that the 
the communication volume is fixed at $O(T\cdot N)$ units, for any number of processors and 
irrespective of the graph density properties.
Furthermore, the partitioning and the communication follow a regular pattern, 
which combined with the simplicity of the scheme, results in minimal implementation overheads.
These factors lead to better scalability.
Finally, the scheme does not require sophisticated hypergraph partitioners and has limited preprocessing cost.

\section{Dynamic GNN Architectures}
\label{sec:models}
We describe the three models used in our experimental study.
They are representative of the dynamic GNN models for DTDG known from 
prior literature (see survey \cite{dynamic-survey}),
making our optimization techniques applicable to the current state of the art.
All the three models follow the framework described in Section \ref{sec:dgnn},
but differ in the choice of the RNN component.

\subsection{{\cdgcn}}
The Concatenate Dynamic GCN \cite{cdgcn} uses the well-known LSTM \cite{lstm} for RNN temporal aggregation.
At a high level, referring Equation \ref{eqn:rnn}, LSTM state $S_t$ consists of a pair $(h_t, c_t)$ referred as 
the hidden and the cell memory. At timestep $t$, the state $S_t$ and the output $Y_t$ are derived
from the previous state $S_{t-1}$, the current input $X_t$ and the previous output $Y_{t-1}$. 
Thus, the LSTM maintains a window length of $w=1$.

Based on accuracy considerations, {\cdgcn} incorporates skip-connection to GCN 
by concatenating the input features to the output, via modifying Equation (\ref{eqn:bbb}):
$$Y_0 = \wt{A}\cdot X,\quad \quad  Y_1 = Y_0\cdot W, \quad\quad Y = \sigma(Y_0\circ Y_1),$$
where $Y_0\circ Y_1$ represents concatenation. As a result, $Y$ will have $F+F'$ features.
The {\cdgcn} as proposed in \cite{cdgcn} comprises of a single dynamic GNN layer given by a GCN-LSTM pair.
We extend this architecture to two layers in the interest of generality of our study.
This will allow similar deeper models, to make use of our acceleration strategies.

\subsection{{\egcn}}
The \emph{evolving} GCN \cite{egcn} model also uses LSTM,
but incorporates two interesting aspects. First, it maintains a different GCN weight matrix $W_t$ for each timestep $t$
so that Equation~\ref{eqn:bbb} is modified as:
\begin{eqnarray*}
%\label{eqn:egcn}
Y_t = \sigma(\wt{A_t}\cdot X_t\cdot W_t).
\end{eqnarray*}
Secondly, instead of applying LSTM over the vertex features of the graph, the model applies LSTM over the weight matrices.
Each layer therefore performs the following operations:
\begin{eqnarray*}
%\label{eqn:egcnh}
W_t = LSTM(W_{t-1}), \\
Y_t = GCN(A_t, X_t, W_t),
\end{eqnarray*}
Intuitively, the weights evolve over the timeline and directly imbibe the temporal properties.
The paper offers two variants namely, EGCN-O and EGCN-H. The above model corresponds to EGCN-O.

\subsection{{\tmgcn}}
In contrast to {\cdgcn} and {\egcn}, 
for the RNN component, the {\tmgcn} model employs M-transform \cite{mproduct},
a parameter-less temporal aggregation mechanism.
Given input features $X_1[u], \ldots, X_T[u]$ for a vertex $u$,
the output sequence $Y_1[u], \ldots, Y_T[u]$ is obtained 
by  aggregating the current  and the previous $w$ input features at each timestep $t$:
\[
Y_t[u] = \agg(X_{t-w}[u], \ldots, X_t[u]),
\]
where $w$ is the  tunable window size and aggregation is weighted averaging.

Equivalently, the M-transform can be expressed in terms of tensor operations.
Let $M$ be a $T\times T$ lower diagonal matrix. 
Given an input tensor $\calX$ of size $T\times N\times F$, the M-transform is given by:
$\calY = \calX \times_1 M$,
where $\times_1$ refers to the first-mode tensor-times-matrix (TTM) product.
The output tensor $\calY$ has same size as $\calX$.
The temporal effect is restricted to the prior $w$ steps by defining $M$ as:
\[
M_{tk} = 
     \begin{cases}
        \frac{1}{min(w,t)} &\quad\text{if} \max(1, t-w+1)\le k \le t,\\
       0 &\quad\text{otherwise.} \\ 
     \end{cases}.
\]
The choice of weights result in averaging and normalizing the features over the timeline.

\subsection{Smoothening the Input Graphs}
Real world dynamic graphs tend to be extremely sparse.
Towards increasing the density and to maintain  continuity over consecutive snapshots, 
{\egcn} and {\tmgcn} smoothen the snapshots via the notion of edge-life and
M-transform, respectively.

The edge-life transformation carries edges from each snapshot to the subsequent $l$ snapshots
by modifying each $A_t$ as:
$$A_t = A_t + \sum_{i=t-l+1}^{t-1} A_i,$$
where the parameter $l$, called edge-life, is a tunable parameter.
The transformation introduces changes into the graph topology at a slower pace
and increases the density as well.

The {\tmgcn} model implements smoothening by applying the M-transform to the input tensor $\calA$,
as well the input feature tensor $\calX$. In practice, the two mechanisms achieve similar smoothing effect
and both are applied in a pre-processing step.

\subsection{Implementation Aspects}
Our implementation of the three models follows a common framework incorporating graph-difference 
and snapshot-partitioning. Below,
we highlight implementation aspects specific to the models.

The {\egcn} model maintains a separate GCN weight matrix $W_t$ for each timestep.
These matrices are small in size and we store copies in each processor.
The model applies the LSTM operation over the above weight matrices, as against the feature matrices.
Consequently, the LSTM operation can be executed by each processor without 
having to communicate with the other processors. Thus, in addition to GCN, the LSTM
component also becomes communication free. To rephrase, each processor acts independently on
on the snapshots assigned to it. The backpropagation is also executed in a similar manner
and partial gradients for the model parameters are derived. At the end of the training epoch,
these gradients are aggregated across the processors via an all-reduce operation.
This constitutes the only communication and the volume is insignificant since the weight matrices
are small in size.
The M-transform based smoothening used in {\tmgcn} is also executed as pre-processing.

For all the three models, we optimize the spatial aggregation of the first GCN layer via pre-computation. The GCN operation (Equation \ref{eqn:bbb})
can be split as $Y' = \wt{A}\cdot X$ and $Y = Y'\cdot W$. Notice that the first part is 
independent of any model parameters. So, we pre-compute the product and reuse the result in 
each training epoch. Since the operation is an expensive sparse-dense mulitplication, 
this pre-processing improves training time for the baseline as well.

\newcommand{\gd} {{\sf GD}}
\newcommand{\base} {{\sf Base}}
\newcommand{\bsizep} {{\tt bsize}_p}

\begin{table}[t]
\centering
\begin{tabular}{c|ccccc}
\textbf{}           & \textbf{N} & \textbf{T} & \textbf{nnz} & \textbf{M-product}    & \textbf{edge-life}\\
\hline
\textbf{epinions}   & 755 K      & 501        & 13 M         & 653 M                    & 1038 M         \\
\textbf{flickr}     & 2.3 M      & 134        & 33 M         & 963 M                    & 796 M           \\
\textbf{youtube}    & 3.2 M      & 203        & 12 M         & 851 M                    & 802 M            \\
\textbf{AMLSim}     & 1 M        & 200        & 124 M        & 1094 M                   & 1038 M       
\end{tabular}
\caption{
Datasets. For each dataset, the number of vertices ($N$), timesteps ($T$), total number of edges or non-zero elements (nnz)
across all the snapshots are shown. {\tmgcn} and the {\egcn} smoothen the input graph by applying M-product and edge-life,
which introduces new non-zero elements. The number of non-zero elements
after each of the operations is given in the last two columns. The two models are trained on the respective smoothened graphs.}
\label{tab:dataset}
\end{table}

\section{Experimental Evaluation}
In this section, we present an experimental evaluation, first focusing on our CPU-GPU optimizations 
and snapshot-partitioning, followed by a preliminary comparison to the vertex-partitioning approach.
While snapshot-partitioning offers better scaling, it has certain limitations when the individual snapshots are large.
We briefly describe possible strategies for addressing them.

\subsection{Setup}
\mypara{System}
The experiments were conducted on the AiMOS system (\url{https://cci.rpi.edu/aimos}).
Our setup uses $16$ nodes, each with $8$ GPUs, leading to a total of $128$ GPUs. 
Each node has 2x20 cores of 2.5GHz Intel Xeon Gold 6248 and has 768 GiB RAM (shared by the $8$ GPUs).
Each GPU is NVIDIA Tesla V100 with 32 GiB HBM. The nodes are connected by Dual 100 Gb EDR Infiniband. 
In each node, we run up to $8$ processes, each controlling a single GPU and  
mapped to a separate core of the node. 
We use PyTorch 1.7.1 for training, NCCL 2.8.4 for backend communication and PyNCCL 0.1.2 for collective routines.
All our codes are implemented in python. 

\mypara{Dataset}
Our benchmark consists of four datasets shown in Table \ref{tab:dataset}.
Epinions is derived from a user-product rating system,
wheres Youtube represents user-user links and Fickr is based on links among images.
The edges for each of these datasets are timestamped with the time at which the links are formed.
All the three datasets were obtained from Networks Repository \cite{dataset}.
AML-Sim is generated from an Anti-money laundering simulator \cite{amlsim}.
The metadata for these datasets is shown in Table~\ref{tab:dataset}.
As discussed earlier (Section \ref{sec:models}), {\tmgcn} and {\egcn} smoothen
the input graphs by applying the M-product and the edge-life operations in a preprocessing
step. The process increases the size (number of edges) of the snapshots. The sizes of the
input and the smoothened graphs are shown in the table. The models are trained
on the respective smoothened graphs. For instance, for the AMLsim graph, {\tmgcn} is trained
on a graph of size $1094 M$ edges. 

\begin{figure}
\centering
\includegraphics[width=6in]{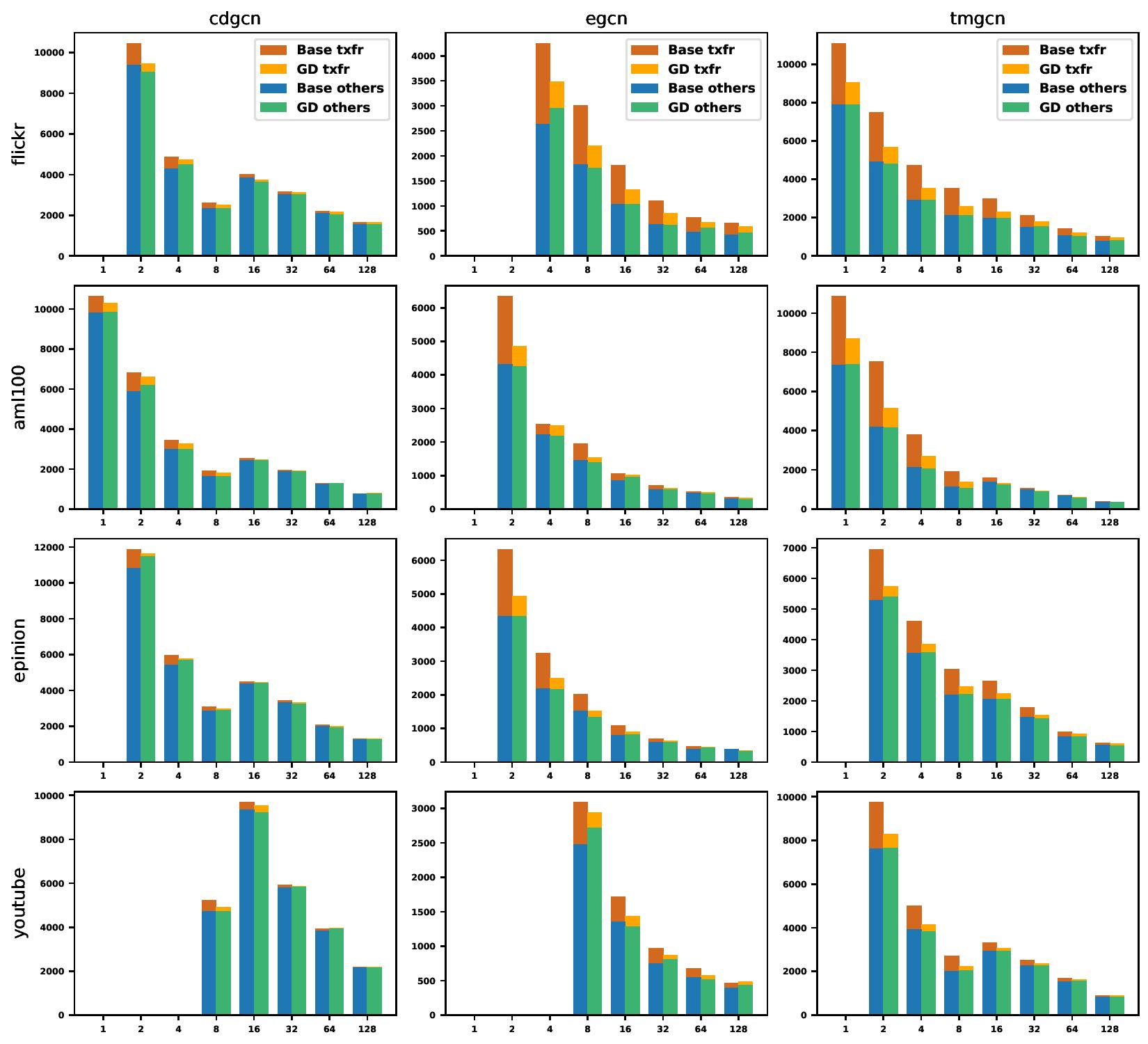}
\caption{
Evaluation of graph-difference technique. Comparison of the naive baseline (Base) and the graph-difference ({\gd})
snapshot transfer methods are shown for each dataset-model pair. In all the plots, X-axis is the number of GPUs and Y-axis is the execution time in milliseconds. Each datapoint is split into two components: the transfer time, and others, which includes the computation and communication time. In some cases, the models did not execute on small number of GPUs due to insufficient memory and  these are left blank.}
\label{fig:expt-delta}
\end{figure}

\mypara{Models and Evaluation}
We evaluate our optimization techniques on three representative  
dynamic GNN models: {\cdgcn} \cite{cdgcn}, {\egcn} \cite{egcn} and {\tmgcn} \cite{tmgcn}.
For all the model-dataset configurations, we use the in and out degrees as the input features, 
as done in {\tmgcn} \cite{tmgcn}.
The intermediate feature lengths are set to $6$ and the number of classification categories is $2$.

As discussed earlier (Section \ref{sec:dgnn}), the dynamic GNN models generate vertex-level embeddings.
Edge-level embeddings can be derived by concatenating the embeddings of $u$ and $v$
for each edge $(u, v)$. 
These embeddings can be used in different ways depending on the task under consideration
such as vertex classification and link prediction.
The first part of our study is concerned
with analyzing the running time performance of our optimization strategies and snapshot-partitioning.
For this purpose, we measure time taken for generating the embedding (and the corresponding backpropagation) 
per training epoch, averaged over $5$ epochs. 
The subsequent segment of the study compares snapshot-partitioning with vertex-partitioning,
which includes an analysis of the loss/accuracy convergence behavior. For this purpose, 
we consider the specific task of link prediction.

\begin{figure}
\centering
\includegraphics[width=6in]{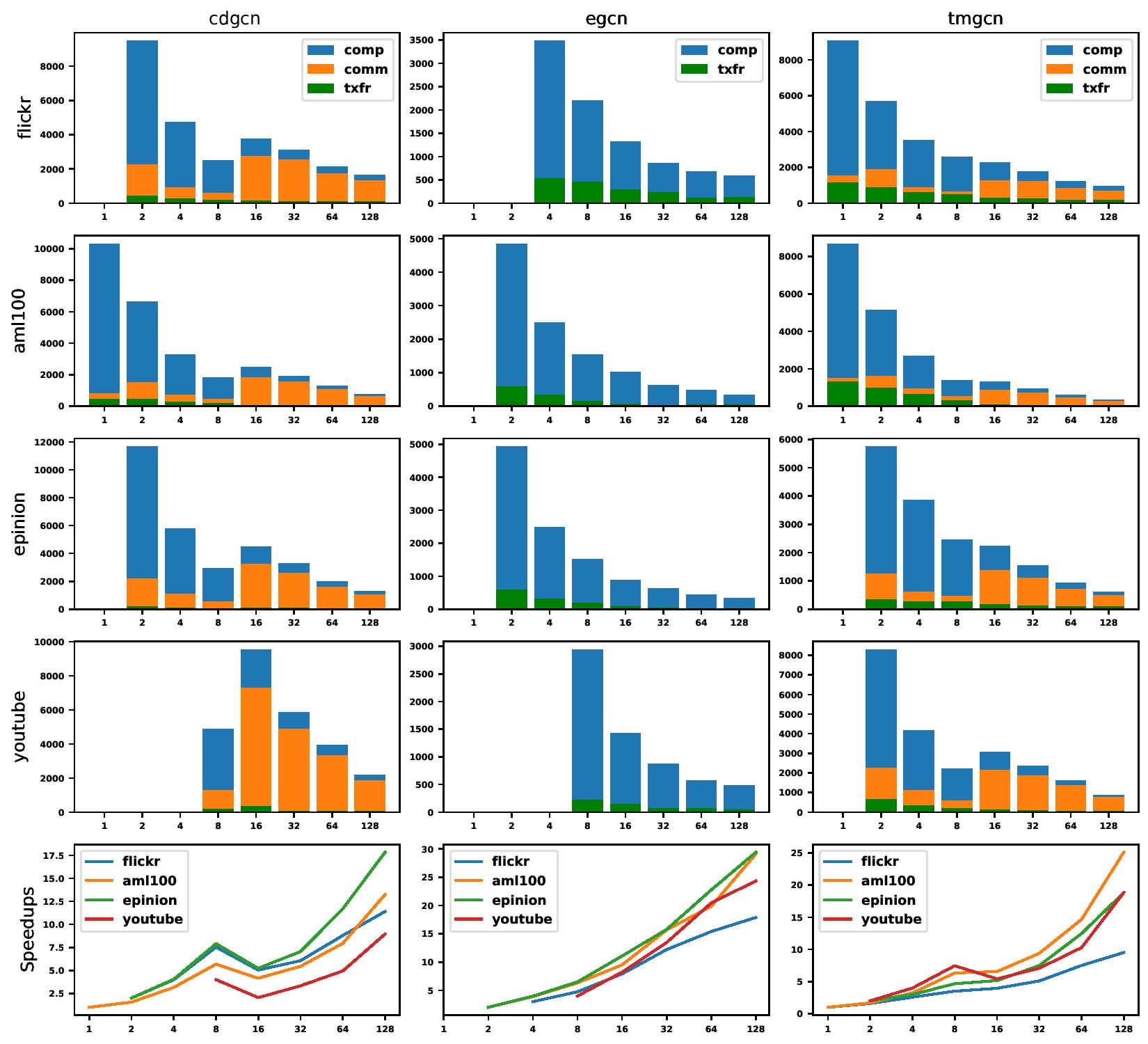}
\caption{
Strong scaling. Results for each  dataset-model pair is shown. 
The implementation is endowed with the {\gd} technique for snapshot transfer.
The X-axis is the number of GPUs and Y-axis is the execution time in milliseconds. 
Each datapoint is split into three components: the transfer time, computation time and the communication time.
For each model, the last plot provides a summary containing the speedup on all the datasets for different values of $P$
with respect to $P=1$ as the reference. 
For configurations where a single processor could not execute due to insufficient
GPU memory, the smallest number of processors $P$ where the execution completed is taken as the reference. 
Since the reference point $P$ varies across different datasets,
for ease of comparison, we take the speedup at $P$ processors as $P$.
}
\label{fig:expt-scaling}
\end{figure}

\begin{figure}
\centering
\includegraphics[width=6in, height=2in]{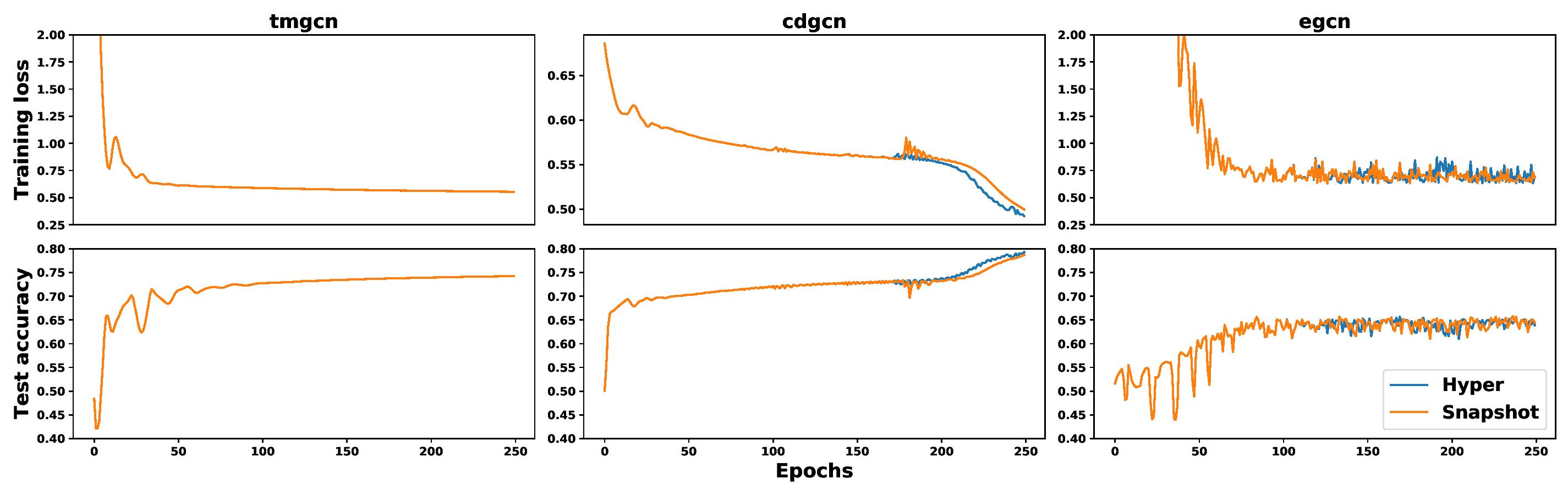}
\caption{
Loss and test accuracy convergence under snapshot and hypergraph partitioning schemes for the three models on the AML-Sim dataset. The curves for {\tmgcn} are identical.}
\label{fig:loss_convergence}
\end{figure}

\subsection{Checkpoint and Graph Difference}
We first evaluate the baseline and the checkpoint based implementations.
Across different model-dataset configurations,
we found that the baseline did not execute on a single node, endowed with $8$ GPUs,
due to GPU memory bottleneck. In contrast, the checkpoint based implementation was able to successfully
run on a single node for all the configuration, with even lesser than $8$ GPUs.

As discussed earlier (Section \ref{sec:checkpoint}),
the checkpoint based implementation needs to transfer 
adjacency matrices $A_{s(b)}, \ldots, A_{e(b)}$ from CPU to GPU while executing a block $b$.
This can be accomplished via naively transferring the matrices in sparse representation given by indices and values.
In contrast, our graph-difference based technique saves execution time by 
transferring only the difference of each snapshot with respect  to the previous snapshot. 
We use pinned memory to optimize the both the methods as it avoids the use of paged memory. 

We denote the naive baseline method as {\base} and the graph difference method as {\gd}.
We evaluate the performance of the two methods on all the dataset-model pairs for number of processors $P=1$ to $128$.
The results are shown in Figure \ref{fig:expt-delta}.
For the ease of comparison, we divide the overall execution time
into two components: (i) the snapshot transfer time; (ii) others, which includes computation and inter-GPU communication.

We can see that for the  {\egcn} and {\tmgcn} models, {\gd} provides significant reduction in the transfer time
across the datasets, with the speedup factors as high as $4.1$x. As a result, the overall execution time
improves by up to $40\%$.  As discussed earlier (Section \ref{sec:models}), based on accuracy considerations,
the two models smoothen the input snapshots by applying the edge-life and the M-product operations to the input snapshots. 
These operations magnify the similarity among consecutive snapshots, enhancing the gains for {\gd}.
In contrast, {\cdgcn} works directly with the input snapshots and the gains in transfer time are up to $2$x.
The latter result demonstrates the strong similarity among the snapshots in real-life, 
which can possibly be exploited in other contexts as well.

We can see that the  gains are higher at smaller GPUs, and this is due to the checkpoint mechanism. 
The checkpoint based implementation executes one block at a time.
The first snapshot of each block is transferred naively and the rest of the snapshots are transferred via the {\gd} method.
Thus, the fraction of the snapshots that benefit from {\gd} is given by $(\bsize-1)/\bsize$, 
where $\bsize$ is the  number of timesteps in each block. 
In the multi-GPU setting, each block is partitioned uniformly, with each processor receiving $\bsizep=\bsizep/P$ snapshots.
The requirement of naively transferring the first snapshot applies within the chunk of snapshots assigned to each processor.
Consequently, the fraction of snapshots that benefit from {\gd} becomes $(\bsizep-1)/\bsizep = (\bsize-P)/\bsize$.
As the number of processors $P$ increases, the benefit ratio decreases. 
Furthermore, communication becomes more dominant at higher system sizes. 
Consequently, the {\gd} technique provides higher gains for smaller number of GPUs.
In summary, the checkpoint and the graph-difference mechanisms allow efficient execution of large datasets on a single node.

\subsection{Scaling Study}
\mypara{Strong Scaling}
We next study the strong scaling behavior of the implementation, endowed with the {\gd} technique for snapshot transfer.
The results are shown Figure \ref{fig:expt-scaling}. As before, the results for each dataset-model pair is presented.
The plots provide breakup of the execution time in terms of three components: snapshot transfer, computation and communication.
Apart from the detailed breakup, for each model, a summary plot is included which presents the speedup curves for all the datasets. 
Taking $P=1$ as the reference point, the plot provides the speedup achieved as we increase the number of processors to $128$.

As discussed in Section \ref{sec:models}, the communication volume {\egcn} is insignificant for {\egcn},
and so, only the other two components are shown.
As $P$ increases, each processor handles lesser number of snapshots  and hence, 
the computation time scales well for all the dataset-model configurations.

In contrast, the communication becomes a bottleneck for {\tmgcn} and {\cdgcn} at higher number of processors.
Under snapshot partitioning, the communication volume is $O(T\cdot N)$, 
irrespective of the number of processors $P$. However, the communication time depends upon the system size.
Each node has $8$ GPUs and so for $P\leq 8$, the communication is intra-node and does not involve interconnection network.
Higher number of processors require inter-node communication and 
as a result, we observe a drop in speedup at $P=16$ compared to $P=8$. 
On further analysis, note that the fraction of intra-node volume is $1/K$ and the inter-node volume is $(K-1)/K$, where $K=P/8$ is the number of nodes.  Thus, the inter-node volume increases with the number of nodes. 
On the other hand, the bisection bandwidth increases with $K$.
The combination of the two aspects determine the communication time and the scaling behavior improves
as $K$ increases. At $P=128$, the speedup is up to $30$x, as against the ideal value of $128$x.

\begin{figure}
\centering
\includegraphics[width=3.5in]{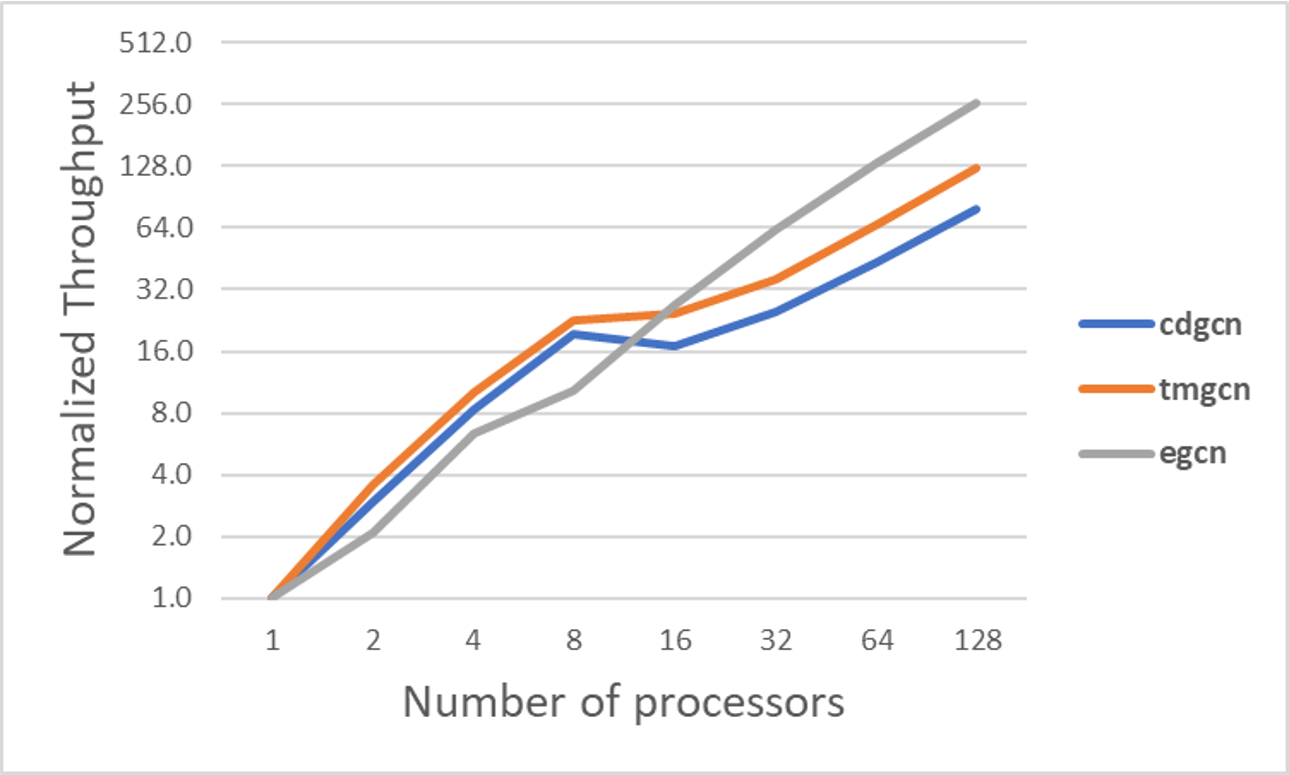}
\caption{Weak scaling of the three models.}
\label{fig:weak}
\end{figure}

\mypara{Weak Scaling}
We next study the weak scaling behavior using randomly generated graphs. Given $T$, $N$ and edge density $f$,
the generator constructs each snapshot independently by adding $N$ vertices and 
randomly selecting $m=N\cdot f$ pairs of vertices as edges. 
The edge-life and the M-product operations are applied to the graphs in the case of {\egcn} and {\tmgcn} models, respectively.

We set the number of timesteps $T=256$ and edge density $f=3$.
Starting with $N=2^{14}$ at $P=1$, 
we scale up to $P=128$ processors via doubling $N$ at each step,
so that the number of vertices is $1M$ at $P=128$. In the case of {\tmgcn},
the aggregate number of edges across the snapshots (after M-product)
varied from $16M$ to $2.1B$ for {\tmgcn} and the other models showed similar trend.

The results are shown in Figure \ref{fig:weak}.
We compute the throughput as the ratio of aggregate number of edges across all the snapshots
to the execution time. We derive the speedup by normalizing the throughput with respect to $P=1$ for each model. 
We can see that {\tmgcn} and {\cdgcn} achieve a speedup of $125$x and $79$x at $P=128$,
as against the ideal value of $128$. The scaling briefly drops going from $P=8$
to $P=16$. The reason is that each node has $8$ GPUs, and hence the node boundary is crossed at $P=16$,
resulting in the use of slower inter-node communication links. The {\egcn} model involves
communication only for gradient aggregation and achieves superlinear speedup of $260$x at $P=128$.

\subsection{Comparison with Vertex-Partitioning}
We present a preliminary empirical comparison of our snapshot-partitioning
scheme with the vertex-distribution method based on hypergraph partitioning (Section \ref{sec:dist}),
illustrating the benefits of snapshot-partitioning discussed therein.
While hypergraph-based partitioning has been well studied in the context of 
graph processing and static GNN, no prior implementation is available for our dynamic GNN setting.
Towards enabling the study, we developed a basic implementation of the strategy.

\mypara{Vertex-Partitioning Implementation}
We use PaToH \cite{patoh} hypergraph partitioner to determine the set of vertices $V_p$
owned by each processor $p$. The vertices $V_p$ need not be consecutive, 
but we make them to be consecutive via renaming, to avoid implementation overheads. 
At each timestep $t$, the $N \times N$ Laplacian sparse matrix $\wt{A}_t$  
and the $N\times F$ feature matrix $X_t$ are distributed by assigning 
the sub-matrices $\wt{A}_t[V_p,:]$ (rows corresponding vertices in $V_p$)
and $X_t[V_p,:]$ to the processor $p$.
Since RNN operates on each vertex independently over the timeline, 
it can be executed via kernel calls without the need for communication.
However, the SpMM convolution operation $Y_t = \wt{A}_t \cdot X_t$ is more involved and requires
communication. We want the result $Y_t$ to be distributed in the same manner so that
$p$ derives $Y_t[V_p,:]$. In this computation, $p$ requires the row $X_t[v,:]$,
for a vertex $v$, only if the corresponding column $\wt{A}_t[:,v]$ contains
at least one non-zero element (alternatively, $p$ owns a neighbor of $v$).
To reduce the communication, any processor sends only the required rows to the other processors.
The hypergraph partitioner is set up in a such a manner that the above communication volume is minimized.
To avoid overheads during training time, the indices are pre-computed 
so that each processor knows the rows it needs to send to every other processor. 
Our implementation ensures that data structures such as the above indices are maintained in-place on the GPU. 

\mypara{Link Prediction}
%As discussed earlier (Section \ref{sec:dgnn}), apart from vertex classification, the models can also be used for analyzing the edges.
%In the latter case, we derive an embedding for an edge $(u, v)$ by concatenating the embeddings of $u$ and $v$.
%As before, by applying a fully-connected layer to these embeddings, 
%we can perform  tasks such as classification and prediction.
For the purpose of comparing the two partitioning schemes, 
we study the link prediction problem considered in prior work on dynamic GNN \cite{tmgcn,egcn}. 
The objective is to  train on the first $T$ timesteps and predict edges that might appear on
timestep $T+1$.  To construct the training set, 
for each timestep, we select $\theta$ fraction of the edges in $G_t$ 
and assign them label $1$, and include an equal number of 
randomly chosen vertex-pairs $(u,v)$, with label $0$.
The testing sample at timestep $T+1$ is constructed in a similar manner
from the graph $G_{T+1}$. The test accuracy is measured as the percentage of 
correctly classified pairs. The parameter $\theta$ controls the size of the
training set and we set it to $0.1$ in our experiments.
The dynamic GNN models produce an embedding for each vertex at each timestep. 
We derive classification for a pair of
vertices $(u,v)$ by concatenating the embedding of the two endpoints  and applying a fully connected layer.

\mypara{Evaluation}
We illustrate the benefits of snapshot partitioning  by considering the AML-Sim dataset. 
We execute all the three models on this dataset under the two partitioning schemes.
We provide the gradient checkpoint mechanism to hypergraph partitioning as well,
to avoid GPU memory bottlenecks. 
Snapshot-partitioning is endowed
with the graph-difference based CPU-GPU transfer of snapshots,
whereas the hypergraph partitioning transfers the snapshots directly.
We execute GCN and RNN as single-batch operations. 
The results of the evaluation are shown in Table \ref{tab:lp} (averaged over five epochs).

\mypara{Loss Convergence}
Before analyzing the execution time performance, we first consider the convergence of loss and accuracy
under the two partition schemes. Unlike deep neural networks, our processing does not involve (variable sized) batched 
gradient descent or batch normalization layers that impact final accuracy.
Consequently, both the schemes simulate the underlying sequential algorithms faithfully.
As a result, their convergence behaviors are identical, except for floating point 
accumulation errors. This is illustrated by Figure \ref{fig:loss_convergence},
which shows the cross-entropy loss and test accuracy for the two schemes.
We can see that the curves are identical under the two schemes for the {\tmgcn} model,
and diverge mildly towards the end for {\cdgcn}.
There is noticeable differences on {\egcn}, 
however the underlying (sequential) loss and accuracy in this case show considerable
fluctuations within consecutive epochs.
Given that the two models simulate the convergence of the sequential model,
we can compare their execution time performance on a per-epoch basis.

\mypara{Communication Volume}
Snapshot-partitioning incurs a volume of  $O(T\cdot N \cdot (P-1)/P)$ (excluding self-communication),
that approaches the fixed limit of $O(T\cdot N)$ units as the number of processors $P$ increases.
In contrast, the  volume under vertex-partitioning grows with $P$, as more edges get split among the processors.
The behavior is illustrated in the table. On the {\tmgcn} model, 
hypergraph-partitioning volume is lesser at $P=4$, nearly matches at $16$ processors, and overshoots at $P=64$. 
The {\egcn} model applies RNN over the locally-held copies of the weight matrices, as against feature matrices.
Hence, snapshot-partitioning is communication free, except for an insignificant gradient aggregation,
and is clearly superior. In contrast, {\cdgcn} does not 
smoothen the input graph  (via M-product or edge-life), resulting in a sparser model-training graph.
The vertex-partitioning volume still increases with $P$, 
but stays lower than that of snapshot-partitioning till $P=64$.

\begin{table}[t]
\centering
\begin{tabular}{l||c||c|c|c|c}
\multirow{2}{*}{\textbf{Model}} & \multirow{2}{*}{\textbf{Ranks}} & \multicolumn{2}{c|}{\textbf{Comm Volume}} & \multicolumn{2}{c}{\textbf{Time (ms)}} \\
               &                & {\em snapshot} & {\em hyper} & {\em snapshot} & {\em hyper} \\
\hline
\hline
\multirow{3}{*}{\textbf{tmgcn}} &  4   &    5.2   &   3.2     &   3396    &    6668    \\
                                & 16   &    6.5   &   6.8     &   1384    &    5254    \\
                                & 64   &    6.8   &   9.5     &    593    &    9164    \\
\hline
\multirow{3}{*}{\textbf{cdgcn}} &  4   &   13.8  &    0.4     &   3867    &    6252    \\
                                & 16   &   17.3  &    0.9     &   2545    &    4653    \\
                                & 64   &   18.1  &    1.2     &   1135    &    8856    \\
\hline
\multirow{3}{*}{\textbf{egcn}}  &  4   &   0     &    DNR     &   4185    &    DNR    \\
                                & 16   &   0     &    5.0     &    944    &    8431   \\
                                & 64   &   0     &    6.9     &    308    &    12276    \\
\hline
\end{tabular}
\caption{
Comparison of snapshot and baseline hypergraph partitioning. Volume in billions of floating point numbers.}
\label{tab:lp}
\end{table}

\mypara{Execution Time}
The communication process under vertex-partitioning involves send-recv buffer constructions,
and maintenance of indices of rows to be communicated between processor pairs. 
The irregular indexing and buffering operations induce significant overheads, especially when performed on GPU. 
In contrast, snapshot-partitioning involves a simpler and regular communication pattern:
the snapshot held by a processor is split into equal sized chunks and communicated to the corresponding owners.
This leads to minimal GPU processing and implementation overheads. In addition, the graph-difference based mechanism
reduces CPU-GPU transfer time, leading to superior scaling compared to hypergraph partitioning. 
We note that it may be possible to reduce the implementation overheads of vertex-partitioning.
However, the increasing communication volume and irregular communication pattern will remain impediments to scaling.

\subsection{Limitations and Possible Improvements}
\mypara{Large Snapshots \& Hybrid Partitioning}
The snapshot partitioning scheme assigns each snapshot in its entirety to a processor,
which may be infeasible when the dataset contains large individual snapshots
that are too big to process on a single GPU.
A related issue is that some processors may be left idle
when the number of snapshots ($T$) is smaller than the number of processors ($P$).

A hybrid partitioning scheme is a possible approach to handle the above scenarios.
The idea is to create groups of processors, and divide the individual snapshots into chunks and distribute them with a group.
Existing static GNN partitioning techniques such as block-wise partitioning \cite{buluc} can be adapted
for intra-group distribution. More generally, a hybrid scheme can be designed by combining the 
above approach with snapshot partitioning. 

To explore the possibility, we experimented by training the 
{\tmgcn} model with two large datasets derived from the AML-Sim generator. 
We trained the model on two GPUs by splitting each snapshot between the two.
The datasets characteristics and test accuracy obtained are shown below;
as with the earlier schemes, the implementation truthfully simulates the 
sequential execution. 
\begin{center}
\begin{tabular}{c|c|c|c|c}
Dataset          &    T    &  nnz   &   size    & accuracy\\
            \hline
AMLSim-Large-1   &   200   &  2.2 B &  44 GB    & 63.8\% \\
AMLSim-Large-2   &   200   &  3.2 B &  64 GB    & 65.8\% \\
\end{tabular}
\end{center}
The above experiment shows that it is possible to design techniques which distribute individual snapshots
among multiple processors for handling large snapshots.

\mypara{Computation-Communication Overlap}
The dynamic model execution in each layer involves four steps:
GCN operation; an all-to-all communication for redistribution; 
the RNN operation; an all-to-all redistribution step that prepares for the next layer.
Our current implementation executes the four steps sequentially.
However, it may be possible to overlap the computation and the communication steps, as outlined below.
In the non-checkpoint version,  each processor $p$ owns $b=T/P$ snapshots.
The processors select one of their snapshots and apply the GCN operation.
Then, they re-distribute the results restricted to the selected snapshots.
The above communication can be overlapped with the GCN operation for the next 
set of snapshots. The third and the fourth steps  can be overlapped in a similar manner. 
In the checkpoint version, the same idea can be utilized, but within each checkpoint block.

\section{Conclusions and Future Work}
We presented, to the best of our knowledge, the first study 
on the scalability aspects of training dynamic GNN models.
With a focus on exploring novel opportunities presented by the temporal aspects of dynamic GNNs,
we designed a graph-difference based technique for minimizing the CPU-to-GPU transfer time
and an efficient distribution scheme based on snapshot partitioning.
We list interesting avenues for future work: 
(i) a hybrid partitioning scheme for handling large snapshots; 
(ii)  exploration of computation-communication overlap; 
(iii) scaling of Continuous Time Dynamic Graphs (CTDG),
wherein the evolving graph is represented by insertion/deletion of vertices/edges. 

\subsubsection*{Acknowledgements} 
We thank the anonymous reviewers of a conference version of the paper for their insightful comments and suggestions that helped in improving the paper considerably.

\bibliographystyle{plain}
\bibliography{main}

\end{document}